\documentstyle[12pt,aaspp4]{article}

\newcommand{\msun}{M_{\odot}}
\newcommand{\lsun}{L_{\odot}}

\newcommand{\kms}{\, {\rm km\, s}^{-1}}

\newcommand{\kpc}{\, {\rm kpc}}
\newcommand{\pc}{\, {\rm pc}}

\newcommand{\etal}{et al.\ }

\newcommand{\degree}{^{\circ}}

\begin{document}

\title{Microlensing in the Galactic Bulge: Effects of the Disk Behind the
Bulge}

\author{Vibhat Nair and Jordi Miralda-Escud\'e$^{1}$}
\affil{  Department of Physics and Astronomy,University of Pennsylvania, 
	Philadelphia, PA 19104}
\authoremail{ nair@student.physics.upenn.edu, jordi@llull.physics.upenn.edu}
\affil{$^{1}$ Alfred P. Sloan Fellow}

\begin{abstract}

  A large number of microlensing events have been observed in the
direction of the Galactic bulge, with a measured optical depth in the
range $2 - 3\times 10^{-6}$. It has been shown that most of these
events are due to bulge stars being lensed by other bulge stars or
by foreground disk stars. Among the stars observed in the bulge fields,
there should also be disk stars located behind the bulge; here, we
consider their effect on the microlensing rates. The optical
depth of background disk stars is much higher than that of typical bulge
stars, reaching $10^{-5}$ at $6\kpc$ behind the bulge. Thus, although
background disk stars are a very small fraction of the stars in Baade's
window, we find that $\sim 5$ to $10\%$ of the optical depth should
be due to disk stars more than $3\kpc$ behind the bulge. This fraction
is sensitive to the luminosity function of disk stars at
large scale-height, to the magnitude cutoff of the survey, and to the
amplification bias effect causing large numbers of ``blended'' events.
We consider also the effect of a warp and flare in the disk at large
distances behind the bulge; this could increase the optical depth from
the background disk to $\sim 20 \%$ of the total. 
Events on background disk stars should on average be longer than other events  
and could be distinguished also by measuring the proper motion or distance of 
the stars that have been microlensed. The number of these events could be
an interesting probe to the structure and stellar population of the
far-side of the Galactic disk.

\end{abstract}
  

\section{Introduction}

  Microlensing has emerged over the last few years as a new field in
astrophysics capable of probing the nature of the dark matter, the
structure of our Galaxy, and the mass distribution of stars (see
Paczy\'nski 1996 for a review). So far,
over 100 events have been detected in the bulge of our Galaxy, near
Baade's window (Alcock \etal 1997a), and over a dozen in the LMC
(Udalski \etal 1993,1994; Alcock \etal 1997b). 
In both locations, the observed optical depth is larger than expected
if we assume that the known stars are the only objects causing the
microlensing events, although there remain substantial
systematic errors in the theoretical calculations due to uncertainties
in the spatial distribution of the stars, as well as statistical errors
in the observational results.

  The estimates of the microlensing optical depth to Baade's window
assumed first that most of the events would be caused by disk stars
lensing sources in
the bulge. After the first few events started to suggest a higher
than expected value of the optical depth, it was pointed out by
Kiraga \& Paczy\'nski (1994) that lenses in the bulge should actually
be more important. For a source in the bulge, an event due to a lensing
star in the bulge should typically have a much smaller Einstein
radius, $\theta_E = [( 4GM/c^2) (D_{ls}/D_l D_s)]^{1/2}$, compared to an 
event due to a disk lens, because the distance ratio $D_{ls}/D_s$ is very
small, given the small size of the bulge compared to its distance from us.
However, the smaller Einstein radii are more than compensated for
by the much larger abundance of bulge stars in the observed fields.
The triaxiality of the bulge turns out to help
as a way to increase the optical depth, due to
the larger separation that is allowed between the lens and the source
if the long axis is at a small angle relative to the line of sight.
Assuming the maximum mass for the bulge that is consistent with
dynamical measurements, an optical depth as high as
$\sim 2\times 10^{-6}$ is possible (Zhao, Spergel, \& Rich 1995).
This is still below the observed value, but consistent with it given
the errorbars
($\tau=2.4 \pm 0.5 \times 10^{-6}$ for all events and 
$\tau=3.9^{+1.8}_{-1.2} \times 10^{-6}$ for clump giant events alone, 
in Alcock \etal 1997a; and $\tau=3.3 \pm 1.2 \times 10^{-6}$ in Udalski 
\etal 1994).

  In this paper, we shall analyze the contribution to the total
optical depth of sources that are at larger distances than the bulge.
These sources should generally belong to the disk of the Milky Way
behind the bulge.
Some of them could also be members of the recently discovered
Sagittarius dwarf (Ibata \etal 1994), which was shown to extend
to fields close to Baade's window from a study of RR Lyrae variables
(Alard 1996). Any such sources should have a much higher optical depth
than the very numerous bulge sources, since they
can be lensed by any bulge star with a large value of the distance
ratio.

  Mollerach \& Roulet (1996) discussed the contribution to the optical
depth from stars in the disk and estimated their event durations.
In their models, they considered only sources within $3 \kpc$ of the
center of the Galaxy. In this paper we shall specifically address the
contribution to the optical depth due to sources at larger distances.
We also study the effect of a perturbation on the shape of the disk
behind the bulge to the optical depth. We find that in general,
the events on background disk stars can only account for a small
fraction of the total optical depth toward the bulge, and therefore
cannot increase the total optical depth significantly above the
predictions obtained by assuming that all sources are bulge stars
(in practice, some sources are also foreground disk stars, and this
reduces the mean optical depth). Nevertheless, the fraction of
background disk events is still large enough to be detectable and
may be of interest for studies of Galactic structure.
     
\section{Models: Distribution of Lenses and Sources}

Our purpose here is to illustrate the probable relative contribution of
the background disk (i.e., the disk behind the bulge) to the optical
depth determined in the microlensing experiments.
We adopt the simple model for the mass
distribution of the bulge derived from the best fit for triaxial models to
the DIRBE maps of the bulge at $2.2 \mu m$ (Model G2 of Dwek \etal 1995).
This consists of a triaxial bulge
with a total mass of $1.8\times 10^{10} \msun$. The density distribution
is as follows:
\begin{eqnarray} 
\rho_{b} &= &2.07 \,\exp \left( - {w^2\over 2} \right) ~\msun \pc^{-3},
\nonumber\\
w^4 &=& \left[ {\left( {x^{'} \over 1580 \pc} \right)^{2} +
\left( { y^{'} \over 620 \pc} \right)^{2}} \right]^{2} +
\left( {z\over 430 \pc} \right)^{4} ~,
\end{eqnarray}
where $x^{'}$ and $y^{'}$ are in the plane of the disk, and the $x^{'}$ 
axis forms an angle of $20^{\circ}$ relative to the x axis, which is
the line from the Sun to the Galactic center (hereafter, GC).
For the disk, we use the fit obtained by Gould, Bahcall, \& Flynn
(1997) to the HST star counts: 

\begin{equation} 
\rho_{d}=~0.055~ \exp\left[ \frac{(R_{0}-R)}{3.5 \kpc} \right] \cdot \\
\left[ 0.28 \cdot \exp\left[ {-|z|\over h_k} \right] +
0.72 \cdot \rm sech^{2} \left[ {|z|\over (2 \cdot h_{n})} \right]
\right] ~\msun \pc^{-3} ~,
\label{vertprof}
\end{equation}
where $R$ is the cylindrical radius from the GC, $R_{0}=8 \kpc$,
and $h_{k}=700~\pc$, $h_n=175 \pc$ are the scale-heights of the thick and
thin disks, respectively. The above model of the disk gives a value of 
$50~\msun / \pc^{2}$ for the total star surface density of the disk at the 
solar radius, which is on the high side of present observational estimates 
of the density of luminous stars, but more consistent with dynamical 
estimates of the disk surface density (Gould 1990).

  To calculate the mean optical depth for a population of sources, one
needs to integrate the number of sources per unit solid angle at each
distance with an apparent magnitude brighter than the limit of the
microlensing experiment. In our models, knowledge of the luminosity
function of the source stars is particularly important, given the wide
range of distances over which sources are located, and given our goal
of comparing the contribution to the total optical depth of stars
from different components of the Galaxy.
We use two models for the luminosity function of stars, consisting of
truncated power-law $\phi(L)\, dL
\propto L^{-\beta-1}\, dL$, with three luminosity intervals
of different $\beta$.
Model 1 is
based on the bulge luminosity function obtained by Holtzman \etal
(1997), and corresponds to an old population of stars:

\begin{eqnarray} 
M_{v} > 4 & :& \beta = -0.25 ~, \nonumber \\
2.75 <  M_{v} < 4 & :& \beta = -3.0 ~,  \\
M_{v} <2.75 & :& \beta = -0.5 ~. \nonumber 
\end{eqnarray}

We assume that both the bulge and the disk have this luminosity
function in Model 1 . Model 2 has the same luminosity function
for bulge stars, but for the disk we use the solar neighborhood luminosity 
function of Wielen \etal (1983), given by :

\begin{eqnarray}
M_{v}> 4 & :&  \beta = -0.25 ~, \nonumber \\
M_{v}< 4 & :&  \beta = -0.6 ~.  
\end{eqnarray}

Notice that the line of sight to Baade's window reaches a large vertical
height at the distance where most of the sources are located,
so Model 1 is probably closer to reality because young stars are
concentrated close to the mid-plane of the disk.
Our results will be displayed for two apparent magnitude cutoffs,
$m_{t}=20$ and $m_{t}=22$.

We also model the possibility that the background disk contains a large
warp and flare. Evans \etal (1997)
suggested that warping and flaring on our side of the disk might affect
the optical depth to the Large Magellanic Cloud, and here we shall
examine if warping and flaring on the opposite side of the disk could
have an important effect on the microlensed stars in the bulge.
Because of the large vertical gradient in the density of stars, 
the number of source stars at large distances behind the bulge
might be altered appreciably by warping and flaring of the magnitude
that is common in other galaxies (see Binney 1992 for a review).
Thus, if the disk bends southward behind the bulge, the number of
disk stars visible in Baade's window (at Galactic latitude
$b= -3.9 \degree$) could be greatly increased.
We choose the following model to illustrate the possible effects:

\begin{eqnarray}
R < R_{0} : & h_{k}=0.7   ~\kpc ~, \nonumber \\
            & h_{n}=0.175 ~\kpc ~, \nonumber \\    
\nonumber \\
R > R_{0} :& h_{k}=0.7   \left( {R - R_{0}\over R_{0} } \right)^2 ~\kpc ~, \\
           & h_{n}=0.175 \left( {R - R_{0}\over R_{0} } \right)^2 ~\kpc ~.
\nonumber 
\end{eqnarray}

The mid-plane of the disk is assumed to be at a height $z_w$ given by
\begin{eqnarray} 
R < R_{0} &:&z_w = 0  ~\kpc , \nonumber \\
R > R_{0} &:& z_w= 1.0 \left( {R-R_{0} \over R_{0}} \right)^2 
\cos( \phi+ 70 \degree ) ~\kpc ~.
\end{eqnarray}

In the flared and warped disk model we use these values for $h_{k}$
and $h_{n}$ in equation \ref{vertprof}, and replace $z$ by $z-z_{w}$.

The above model for the warp and flare of the disk is consistent with 
observations of the distribution of neutral hydrogen in the disk (Diplas 
\& Savage 1990). The observations constrain the angle of the line of
nodes with respect to the Sun-GC line to be small, but with an error
of $\sim 10\degree$, and a larger uncertainty in the far side of the
disk due to the scarcity of observations in that region. Of course,
the larger this angle, the more important the effect of the warp will
be for the microlensing event rates from background disk sources. Here
we choose this angle to be $20\degree$. The
line of nodes does not necessarily have to be at a constant angle,
since the warp could be twisted (i.e., the shape of the disk may in
general have some two-dimensional Gaussian curvature).

\section{Tidal Perturbation of the Disk by the Sagittarius dwarf}

  One of the reasons why the disk could be warped is the gravitational
tidal perturbation caused by a satellite galaxy.
In addition to the above simple model for a warp and flare, we shall
examine another model based on the distortion of the
disk induced by the gravitational perturbation of the Sagittarius dwarf.

  The Sagittarius dwarf, discovered by Ibata \etal (1994, 1995) in a
spectroscopic study of the Galactic bulge, is the closest galaxy to the
Milky Way. It is located $\sim 15 \kpc$ behind the bulge, centered
approximately at $l = 5.6\degree$, $b=-14\degree$, and is extended
perpendicular to the Galactic plane over at least $20\degree$ in the
sky. Its projected shape is elongated with an axis ratio $3:1$. Its
radial velocity is $140 \kms$, moving away from us, and the proper
motion indicates that it is moving upwards toward the disk, although
the component parallel to the disk is not yet accurately measured
(Ibata \etal 1997). The luminosity of the galaxy is $\sim 10^7 \lsun$.

  The tidal perturbation that the Sagittarius dwarf will exert on the
disk depends of course on its past trajectory and on its mass, which is
highly uncertain depending on the mass of the dark matter halo
surrounding the visible stars.
We now describe a simple model that we use to obtain an estimate of the
maximum plausible effect of this tidal perturbation.

  The past trajectory of the Sagittarius dwarf is
calculated as if it was a point mass moving in the Galactic potential,
starting from its present position and velocity. 
Table 5 gives the present velocity of the dwarf in cylindrical coordinates,
relative to the Sun and to the GC,
denoted by subscripts h (for heliocentric) and G respectively.
We fix the present position of the dwarf to be $l=5.6^\circ$, $b=-14^\circ$,
$r_{h}=24 \kpc$, $r_{h}$ being the distance of the dwarf from 
the Sun (Ibata et al.\ 1997). We take $v_{h,r}=140 \kms$ and 
$v_{h,z}=150 \kms$. 
The z velocity is chosen to be small within the range consistent with the 
available error bars (Ibata et al.\ 1997), in order to maximize the
perturbation on the disk. For $v_{G,\phi}$, the transverse
velocity of the dwarf parallel to the disk, we use a range of values between
$ 50 \kms $ and $ 250 \kms $. Because the major axis 
of the dwarf is almost perpendicular to the disk, and the tidal elongation of
the structure of the dwarf is most likely to occur along its orbit,
$v_{G,\phi}$ (which as yet is observationally undetermined) is likely to be 
small. However, the present shape of the dwarf should also depend on its
internal kinematic structure, and a larger azimuthal velocity has not
been ruled out.  Note that we have assumed that the velocity 
of the dwarf is in the same sense as that of the rotation of the disk,
which again maximizes the perturbation on the disk. 

  For the Galactic potential, we use a three-component model with an
exponential disk, a bulge and a halo. We model the bulge as a
$1.8\times 10^{10} \msun$ point mass (sufficient for our purpose since
we only need to evaluate the force at large distances from the GC),
a halo with an 8 \kpc core and velocity dispersion of
$155 \kms$,
and an exponential disk with scale-length $R_d=3.5 \kpc$
and central surface density $\Sigma_0 = 680 ~ M_{\odot}/\pc^{2}$.
The circular velocity profile obtained for this model is nearly
flat from $R=5 \kpc$ to $R=50 \kpc $ at a value of $200 \kms$.

  The trajectories obtained for the different choices of $v_{G,\phi}$
are shown in Figures 1 and 2.

  Given the trajectory of the dwarf, we then
consider stars moving initially in the plane of the disk on circular
orbits, and calculate the perturbation of their trajectories due to the
tidal acceleration caused by the gravity of the Sagittarius dwarf,
equal to the gravitational acceleration acting on the star minus the
acceleration acting on the GC. We use the approximation
that the
gravitational force of the Galaxy is always directed to the center,
even though the star will move out of the plane of the disk as a result
of the tidal force. This is
a reasonable approximation because, locally, the entire disk should
be distorted by the perturbing tidal force in the same way, so the
star should remain in the midplane of the perturbed disk.
On a large scale, the
disk will of course no longer be confined to a plane, but including the
potential of the distorted disk would require a full N-body calculation.
We have instead used this simple approximation, which should be valid
at large radius, where the stars do not complete many orbits over the
duration of the tidal perturbation.

  For the mass density profile of the dwarf, we assume a spherical
singular isothermal halo with a cutoff at 10 \kpc, and a total mass of
$5\times 10^9\msun$. Thus, the mass within 2 \kpc is $10^{9} M_{\odot}$,
in agreement with the model of Ibata \etal (1997) reproducing the
observed stellar velocity dispersion. The extent of the dwarf halo (and
therefore its mass) is generally estimated to be smaller if the dwarf
has already completed several orbits around the Galaxy, because it should
have been tidally disrupted (e.g., Vel\'azquez \& White 1995, Ibata \etal
1997). However, the dwarf could be on its first orbit if its direction
of motion was not the same as the direction of its elongation, or if its
orbit had been recently perturbed. For example, the Sagittarius dwarf
might originally have been a distant satellite of the Magellanic Clouds,
and could be on its first orbit since being tidally pulled from that
system. 
In a recent paper, Zhao (1998a) describes such a collision scenario of the 
Sagittarius dwarf with the Magellanic clouds. While his model assumes
the mass of the Sagittarius dwarf to be much smaller than that of the
Magellanic clouds, the orbit of the dwarf is not constrained well
enough currently for this to be the only possibility. In this paper, we 
seek to estimate the maximum possible effect that the
Sagittarius dwarf might have on the optical depth towards Baade's window.
Towards this end we have chosen the mass of the dwarf to be considerably 
larger than the estimates derived from the current models of its orbit. 
 
  The integration of the orbits of stars in the disk was started at
$3 \times 10^{8}$ years into the past; all stars are initialized on
circular orbits in the disk plane.
The final vertical positions of the stars along our line of sight
through Baade's window are plotted in Figure 3, for the three values
of the present azimuthal velocity of the Sagittarius dwarf.
The vertical displacement in the GC vanishes by definition, since the
tidal force acting on every star is defined relative to the GC. We have
also fixed the displacement to zero at the position of the Sun, by
assuming that the plane of the Galaxy is redefined to contain the
Sun-GC line.

  The effect of the perturbation can be separated into four regions
along the line of sight: (1) The foreground disk stars are displaced
southward relative to the Sun-GC line. (2) Immediately behind the bulge,
stars are displaced northward due to the greater force that has acted
on the GC compared to the Sun or the stars in this location.
(3) Stars lying between $\sim 5$ and $15 \kpc$ behind the bulge are
again displaced southward when the azimuthal velocity of the Sagittarius 
dwarf is high, owing to the more recent force acting on these stars as the
Sagittarius dwarf approaches them from the South, following their
orbital azimuthal motion. This part of the background disk can be pulled
down by as much as $250$ pc, a distance greater than the thin disk
scale-height of 175 pc. (4) At distances greater than $\sim 25 \kpc$,
stars are displaced northward by several scale-heights; this is mainly
due to the tidal force that has pulled down the GC relative to the Sun.
 
  To summarize, the results in Figure 3 show that the only region
where the number of sources could
be increased substantially by this disk perturbation is in the 5 or 10
$\kpc$ immediately preceding the Sagittarius dwarf,
along the line of sight near Baade's window.

 
\section{Results for the Optical Depth: Models with Unperturbed Disk}

  The optical depth of a source as a function of its distance from the
Sun along Baade's window (specifically, along the Galactic coordinates
$\ell = 1 \degree$ and $b=-3.9 \degree$) is shown in Figure 4 for our
unperturbed disk model (the difference of this curve in the models with
a perturbed disk is negligible).
The optical depth increases dramatically at the distance of the bulge,
since most of the lenses are located there. Behind the bulge the optical
depth continues to grow owing to the increasing distance ratio with the
bulge lenses. Because the number of lenses beyond the bulge is very
small, the optical depth approaches a constant value of almost
$2\times 10^{-5}$ at very large distances, nearly ten times greater than
the mean optical depth determined observationally. Thus, even a small
fraction of background disk stars among the observed sources could have
significant effects on the observed microlensing events. A similar
phenomenon might take place for microlensing on the Large Magellanic
Cloud (LMC), if some of the observed stars are in star-forming regions
along tidal debris of the Magellanic clouds located far behind the 
LMC (Zhao 1998b).

  Tables 1 to 4 give the contribution to the number of sources and the
optical depth from various components of our Galaxy. The quantity
$f_{i}$ is the fraction of all the sources brighter than the apparent
magnitude threshold, $m$, which belong to each component $i$. The
mean optical depth $\tau_i$ of each component is
\begin{equation}
\tau_{i}=\frac{\int \rho_{i}(D_{s},m) \,
\tau(D_{s}) \, D_{s}^{2}\, dD_{s}}
{\int \rho_{i}(D_{s},m) \,  D_{s}^{2}\, dD_{s}} \, ,
\end{equation}
where $\rho_{i}(D_{s},m)$ is the density of sources of component
$i$ at distance $D_{s}$ above the magnitude threshold $m$,
and $\tau(D_{s})$ is the optical depth for a source at distance $D_{s}$.
    
  Thus, $\sum ~\tau_{i} f_{i} = \tau$,
where $\tau$ is the mean optical depth for all sources above the
magnitude cutoff. 
 We shall be ignoring dust obscuration here; in
practice, most of the dust in the line of sight to the bulge fields
searched for microlensing is close to the Sun, so practically all the
sources have the same obscuration, and the effect of dust is then to
simply change the value of the apparent magnitude cutoff.

  For both models of the luminosity function, the contribution of the 
background disk to the mean optical depth, usually neglected, is
significant. For the unperturbed disk (see Tables 1 through 4), the
background disk contribution is at least $\sim 14 \%$, and could be as
high as $\sim 27 \%$ in Model 2 for the luminosity function. Notice,
though, that a lot of this contribution is due to sources close to the
center, where the disk should probably be truncated due to the presence
of the bar: the optical depth contribution from background disk stars
located more than 3 \kpc behind the GC is only $\sim 6\%$ in Model 1.
These contributions depend sensitively on the 
luminosity function of the source stars and on the apparent magnitude
cutoff of the survey, as we see in the Tables. 

  In Model 1, the disk has a luminosity function appropriate for old
stars, implying that the number of stars declines abruptly at
luminosities greater than the main-sequence turnoff (eq.\ 3). This
results in a rapid decrease of sources at distances further than the
point where the main-sequence turnoff coincides with the magnitude
limit of the microlensing survey. In the absence of extinction,
this distance is 16 (28) $\kpc$ for a threshold $m_t=20$ (22). This is
the reason why the contribution from stars in the background disk
increases as fainter sources can be observed (or in regions of lower
extinction). Model 2 also predicts a greater fraction of the optical
depth from distant disk stars, owing to the greater abundance of
luminous stars acting as microlensing sources when a young population
is assumed to be present.

  The fraction of the optical depth contributed by sources at each
distance $D_s$ is shown in Figure 5, for
the two luminosity function models for $m_{t}=20$. 
The average duration of events on background disk sources is 
larger than for events on bulge sources (this will be discussed in 
detail toward the end of this section). Because of the longer average
duration of events, the contribution to the rate of events by background
disk sources is smaller than the contribution to the optical depth.

  The results presented in Tables 1-4 and Figure 5 do not take into
account the effect of magnification bias. The fraction of microlensed
stars located at each distance is valid for sources above a fixed
magnitude cutoff when there is no lensing amplification. In practice, 
many of the
microlensing events occur on sources which are fainter than the
magnitude limit of the survey, but are brought above the limit during
the microlensing event. Often these sources will be blended with a
brighter star, so it is not always possible to know what the magnitude
of the unlensed star is. This blending effect should result in an
increased contribution by distant sources to the observed microlensing
events, because the magnitude threshold is effectively changed to
fainter levels. On the other hand, extinction will raise the magnitude
threshold. In order to compare model predictions of the distribution of
source distances with observations, it will be necessary to securely
identify and to measure the unlensed flux of every star that has been
microlensed. Spectroscopy is probably difficult in
these faint stars in crowded fields; but possibly, some clear photometric
indicator that can select stars in the background disk may be identified
(for example, stars located between the giant branch and main sequence
of the bulge stars, after correction for extinction, would probably be
giants in the background disk),
which could make it possible to measure the enhancement of the fraction
of distant sources among the microlensed stars compared to a random
sample of sources in the bulge fields. Another way to distinguish
background disk stars from bulge stars are proper motions.
Background disk stars should always be moving 
along the plane in the direction of decreasing longitude, with 
$\mu \sim (80 {\rm kpc})/D_{s} {\rm mas/yr} $, where 
$D_{s}$ is the distance to the star
from the sun. Thus for a typical disk source behind the bulge at 
$\sim 12 \kpc$, $\mu \sim 6  {\rm mas/yr}$.
Although bulge stars can also 
have similar proper motions, the two populations could be statistically 
separated.

  The inclusion of source stars in the disk for the computation of the
total microlensing optical depth in Baade's window does not change
the result by a large factor, compared to models where only bulge
sources are included. Stars in the background disk certainly increase
the total optical depth, by $\sim 10\%$ depending on the model;
however, foreground disk stars also need to be included as sources,
resulting in a decrease of the optical depth that tends to cancel
the effect of the background disk in most cases. This is the reason
why the total optical depth is generally not very different from the
bulge optical depth.

  If microlensing events could be detected along lines of sight at 
lower Galactic latitude compared to Baade's window, a much larger
contribution to the total optical depth from sources in the background
disk should be expected. As an example, Table 5 gives 
the contributions of various components to the total optical depth for
the line of sight at $l=1 \degree,b=-1 \degree$. In this case, stars 
in the disk beyond 3 $\kpc$ from the bulge account for 5\% of all 
sources and $20 \%$ of the optical depth, compared to 1.6\% and
5\%, respectively, in Baade's window. Observations at these low
Galactic latitudes would only be feasible for a microlensing survey
in the infrared, due to the high obscuration.

  In general, events on background disk stars should be of longer
duration than events on bulge stars, owing to two reasons. First, the
Einstein radius is larger for a source in the disk, due to the larger
distance ratio $D_{ls}/D_{s}$ when the source is in the background disk,
since most of the lenses are in the bulge. Second, the relative proper
motion of the lens and the source is smaller for a disk source. To see
this, consider the proper motions of the lens and the source relative to
the Galactic center. The proper motion of the lens always has the same
distribution, but the proper motion of the source relative to the
Galactic center is generally smaller for a background disk source
because the effect of the velocities of the source and the Sun have 
opposite sign (for example, the proper motion relative to the Galactic
center is zero for a source at the same distance behind the Galactic
center as the Sun). The velocity dispersion of disk sources is also
small compared to that of bulge sources. Therefore, among the longest
events, the fraction of microlensed stars belonging to the background
disk should be higher than for shorter events. 

  Figure 6 shows the distribution of event durations for sources at
three different distances, as indicated in the figure. We assume that
all the lenses have a mass of $1\msun$, for the purpose of illustration.
We also consider only bulge lenses, with a spherically symmetric
Gaussian distribution of velocities with dispersion equal to $150 \kms$.
The disk sources are moving with constant circular velocity
$V_c=200\kms$, and dispersion of $50 \kms$. The calculation was made
using Monte-Carlo simulations of $10^{6}$ events for each source
distance. Sources at $16\kpc$ have longer durations by as much as a
factor 2 compared to sources at 10\kpc (a typical distance for a bulge
source). Introducing a distribution for the mass of the lens will have
the effect of broadening the curves shown in Figure 6. This suggests
that a long event duration may not be a very good discriminant for
background disk events, given the large dispersion in lens mass. At the
same time the contribution of background disk sources to the number of
microlensing events should not be very different from the contribution
to the total optical depth computed in this paper, since the difference
in the mean event duration is not very large.

\section{Effect of a Perturbed Disk on the Optical Depth}

  The flared and warped model of the disk, introduced in \S 2, increases
the number of sources at large distances behind the bulge by reducing
the distance from Baade's window to the plane of the
perturbed disk, and by increasing the disk scale-height. Thus, in Model
1, the disk perturbation
increases the fraction of sources at distances greater than $ 15 \kpc$
by a factor of 1.5, from $0.2\%$ to $0.3\%$, at $m_t=20$. The fraction of
fainter sources $(m_t=22)$ is increased by a factor of 2, to
$1.3\%$, and for Model 2 of the luminosity function this fraction can be
as high as $2.3\%$.

  The flared and warped disk results in a similar increase of the optical
depth due to background
disk stars. For example, in Model 2 and for $m_t=22$, the fraction of the
optical depth from sources at distances greater than $15 \kpc$ is
increased from $5\%$ to $12 \%$.
Figure 5 also shows the
distance distribution of the microlensed stars for the perturbed disk
model, with Model 1 of the luminosity function. 

  These results show how the measurement of the distance
distribution of the microlensed stars in the bulge fields
(which, as mentioned above, could be estimated from photometry and
proper motions) could be used
to constrain the structure of the disk in the far side of the
Galaxy. These constraints would require an accurate knowledge of the
luminosity function of the background disk stars, to which our results
are also very sensitive to. 

\section{Effect of the Sagittarius Dwarf}

  Results are also given in Tables 1 and 2 for the models where the
perturbation of the disk is caused by the tidal force of the
Sagittarius dwarf. The effect of this perturbation is generally very
small. In fact, in many cases the optical depth contributed by the
background disk is decreased, owing to the northward displacement of
the background disk sources close to the bulge (see Figure 3). The
perturbation by the Sagittarius dwarf also causes the foreground disk
(in front of the bulge) to be bent southward. This has two opposite
effects on the total optical depth: the number of disk lenses
increases, so the optical depth of every component also increases; but
the number of foreground disk sources is also increased relative to the
bulge sources, and this reduces the mean optical depth.

  However, the effect of the stars in the Sagittarius dwarf acting as
{\it sources} for microlensing may be important. A simple estimate of
their effect compared to background disk stars at a similar distance
may be obtained as follows. The Sagittarius dwarf has a total
luminosity of $10^7 \lsun$, spread over $\sim 200$ square degrees of the
sky (Ibata \etal 1997), or $30 \kpc^2$. This yields a surface luminosity
density of $0.3 \lsun\pc^{-2}$. For the disk, the surface luminosity
density at the solar orbit is $15\lsun\pc^{-2}$; the Sagittarius dwarf
is about two scale-lengths further away from the center than the Sun,
so the disk surface luminosity density at that distance is
$\sim 1.8 \lsun \pc^{-2}$. Along Baade's window and at
the distance of the Sagittarius dwarf, the height below the plane is
$1.6 \kpc$. Therefore, the luminosity density at
this position integrated over a scale-length,
for our model in equation (\ref{vertprof}),
is reduced by a factor
\begin{equation}
0.72 e^{-9.2}(3.5\kpc/0.175\kpc)/2 +
0.28 e^{-2.3}(3.5\kpc/0.7\kpc)/2 ~, 
\end{equation}
to $0.15 \lsun \pc^2$. According to this,
the Sagittarius dwarf stars are two times as abundant as background
disk stars at the same distance range, although disk stars slightly
closer to us (at distances of one to three scale-lengths behind the
bulge) should dominate over the Sagittarius dwarf stars.
Microlensing events on stars in the background disk and the Sagittarius
dwarf can be distinguished if proper motions are available. Proper
motions have been measured in some stars in Sagittarius dwarf, although
with relatively poor accuracy (Ibata \etal 1997).
 
\section{Conclusions}

  This paper has discussed the contribution of background disk sources
to the total optical depth of microlensing events observed in Baade's window.
This contribution is appreciable and needs to be taken into account in
comparisons between theoretical predictions and the observations.
About $15 \%$ of the optical depth in Baade's window is due to
background disk stars; about half of this is contributed by stars
at a distance greater than $3\kpc$ behind the bulge. This fraction is
sensitive to the luminosity function of the source stars and the
apparent magnitude limit of the
microlensing survey (and therefore, to the ``blending effects''). 
The possibility that the far-side of the disk is strongly warped and
flared has also been considered, and we showed that this could
significantly increase the number of background disk stars in Baade's
window. 
Our results may be taken as upper limits on the contribution of disk 
perturbations to the optical depth, since we have chosen the most 
favorable parameters in our models (e.g., the angle of the line 
of nodes for the warp and flare model was chosen to be $20\degree$,
and a large mass of the Sagittarius dwarf was assumed).  

  Future microlensing surveys towards the bulge could provide an
interesting probe to the structure and the stellar population of the
Galactic disk behind the bulge. We have focused in this paper on
predicting the contribution to the microlensing event rate on Baade's
window, at a Galactic latitude $b\simeq -4\degree$, because the present
surveys are being done near this area of low extinction. However,
in lines of sight closer to the Galactic plane, the density of disk
stars should decrease more slowly with distance, and the contribution
of background disk stars should therefore be much higher; an example
has been presented in Table 5. Extinction
should of course be very high at lower Galactic latitude, so
microlensing surveys in this area would probably need to be carried out
in the infrared. This may become more feasible as the size of CCD
cameras sensitive in the infrared increases.
  
We would like to thank the referee, Dr. HongSheng Zhao, for his comments and 
suggestions that have improved the content and the presentation of this paper.

\newpage
\begin{figure}
\centerline{
\hbox{
\epsfxsize=4.4truein
\epsfbox[55 32 525 706]{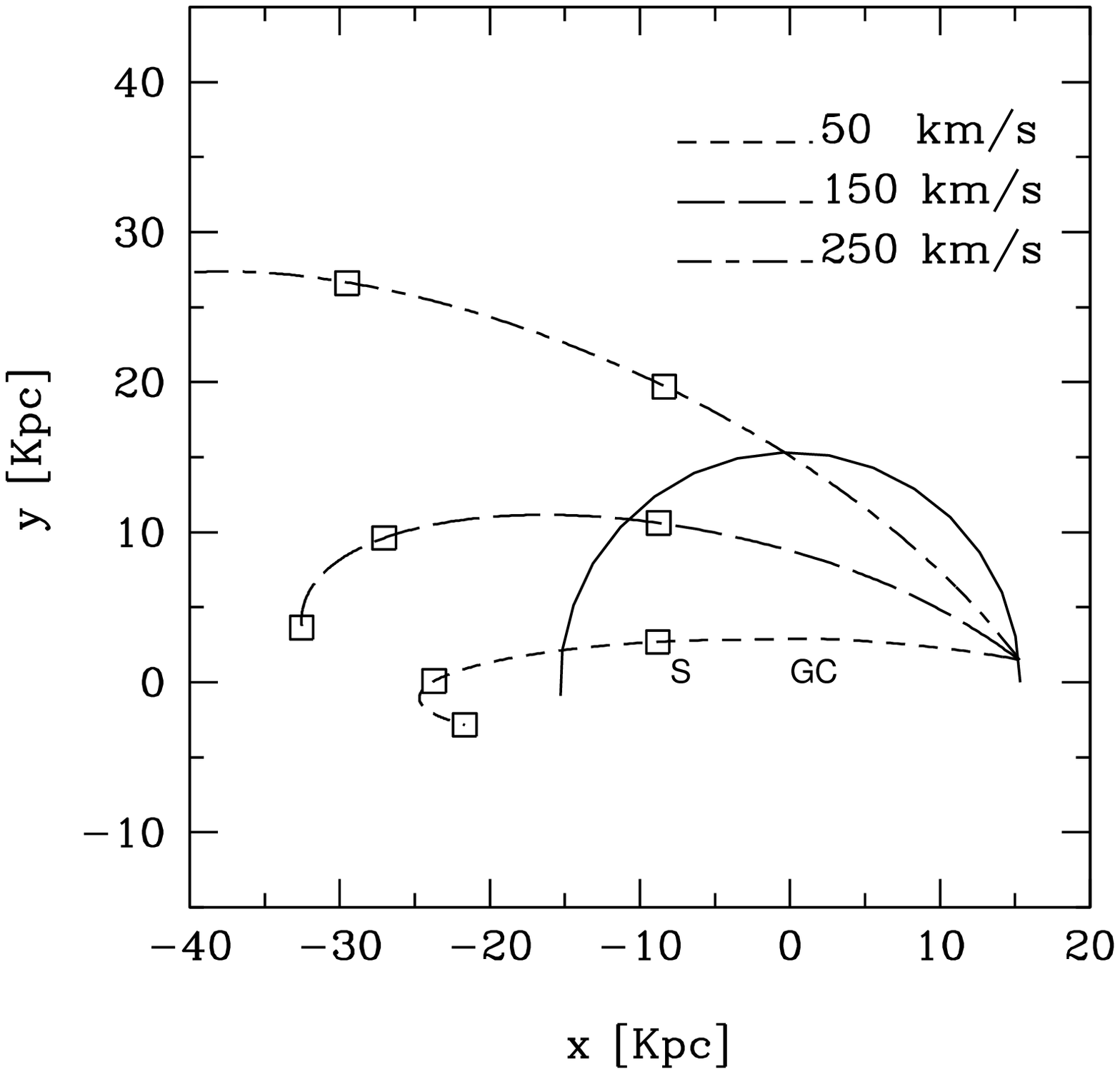}
}
}
\vskip -40pt
\caption{
Orbit of the Sagittarius dwarf projected on the x-y plane, for the three
cases of $v_{G,\phi}$ indicated in the figure. The positions at 
$10^{8}$ year intervals are marked by squares. The circle has been shown to 
indicate the unperturbed orbit of a star at a radius of 15 \kpc. 
}
\end{figure}
\vfill\eject

\begin{figure}
\centerline{
\hbox{
\epsfxsize=4.4truein
\epsfbox[55 32 525 706]{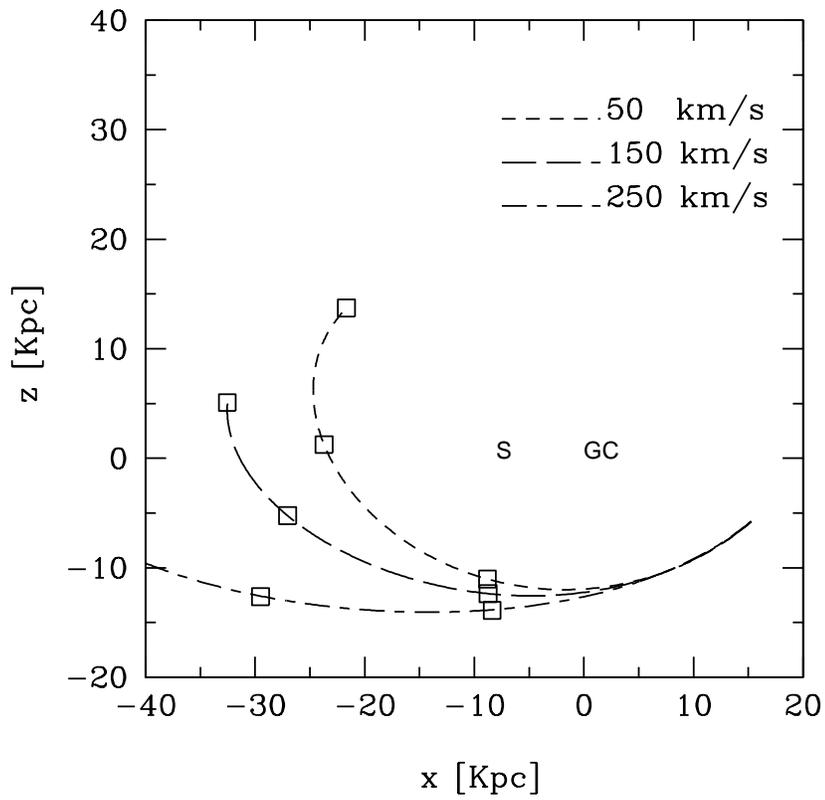}
}
}
\vskip -40pt
\caption{Orbit of the Sagittarius dwarf projected on the x-z plane, for the 
three cases of $v_{G,\phi}$ indicated in the figure. The positions at 
$10^{8}$ year intervals are marked by squares.
}
\end{figure}
\vfill\eject

\begin{figure}
\centerline{
\hbox{
\epsfxsize=4.4truein
\epsfbox[55 32 525 706]{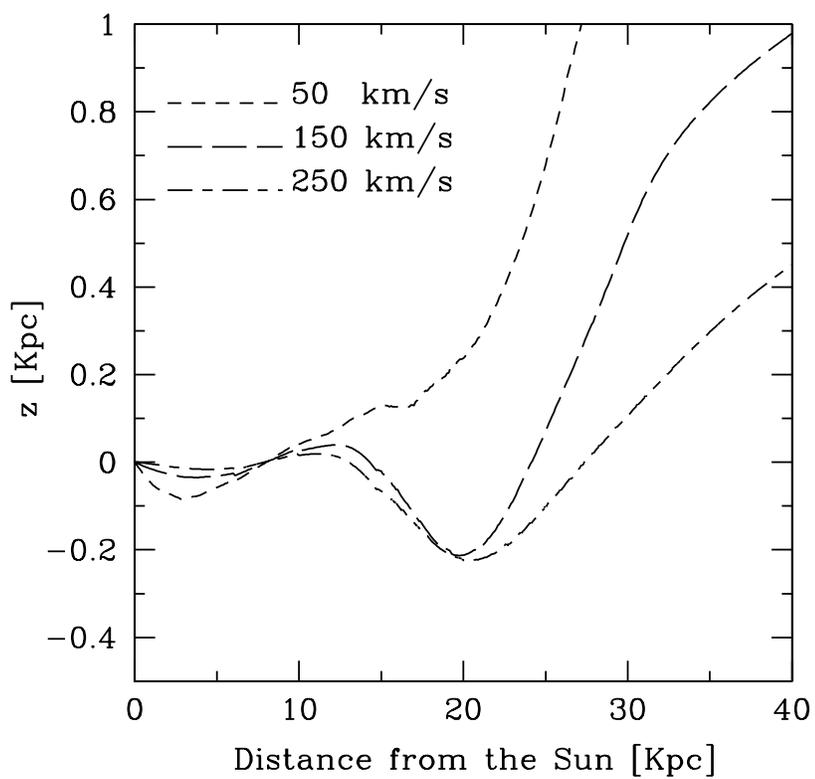}
}
}
\vskip -40pt
\caption{Final height of the stars with respect to the sun-GC line, for the
three cases of $v_{G,\phi}$ indicated in the figure. 
}
\end{figure}
\vfill\eject

\begin{figure}
\centerline{
\hbox{
\epsfxsize=4.4truein
\epsfbox[55 32 525 706]{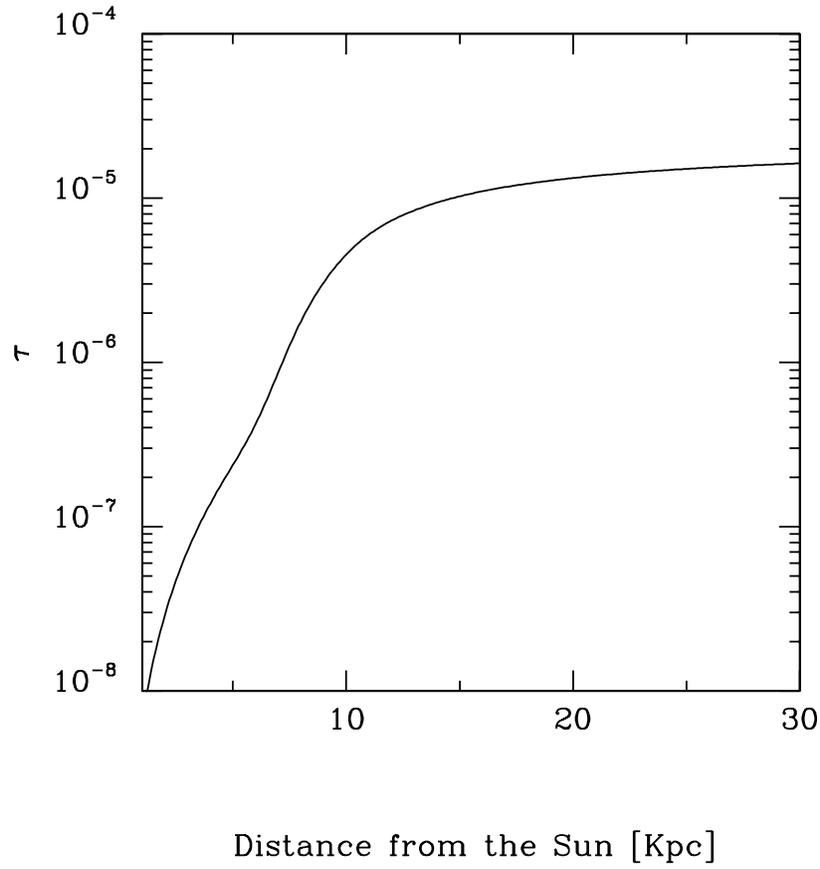}
}
}
\vskip -40pt
\caption{The optical depth $\tau$, as a function of source distance along 
our line of sight through Baade's window. Only the unperturbed case is 
shown. The difference from the unperturbed case is negligible for other 
cases. 
}
\end{figure}
\vfill\eject

\begin{figure}
\centerline{
\hbox{
\epsfxsize=4.4truein
\epsfbox[55 32 525 706]{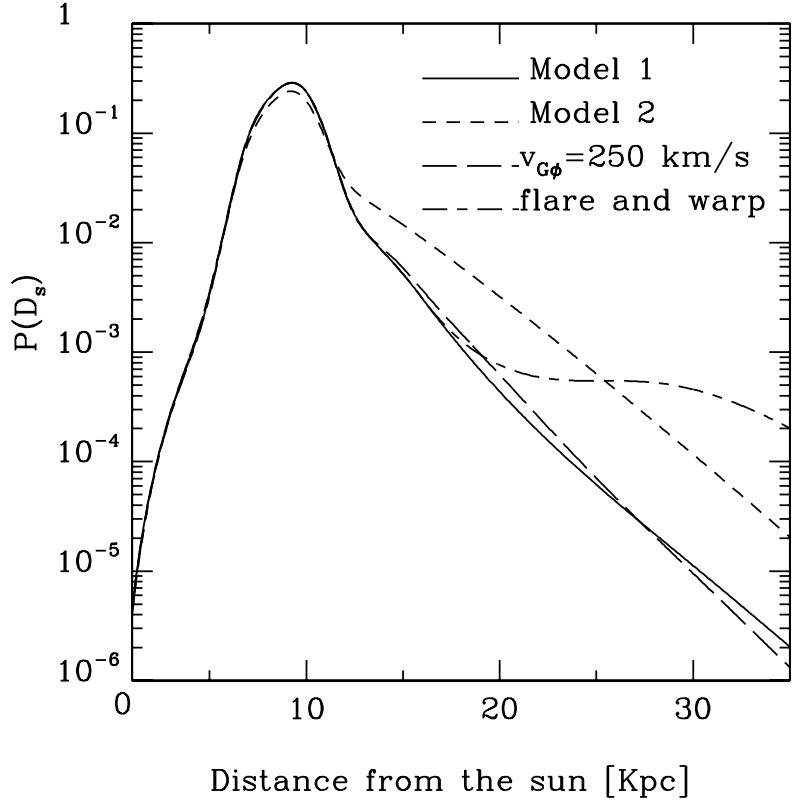}
}
}
\vskip -40pt
\caption{$P(D_{s}) \cdot dD_{s}$ is the fraction of the optical depth 
contributed by sources at distances between $D_{s}$ and $D_{s}+dD_{s}$ from 
the sun. The solid, long dashed and dot dashed line correspond to an 
unperturbed disk, disk perturbed by the Sagittarius dwarf galaxy with 
$v_{G,\phi}= 250 \kms$, and a warped and flared model of the disk, 
respectively. For all three we assume Model 1 for the luminosity 
function and a magnitude cutoff of 20. The short-dashed line corresponds
to an unperturbed disk with Model 2 for the luminosity function and a 
magnitude cutoff of 20.
}
\end{figure}
\vfill\eject

\begin{figure}
\centerline{
\hbox{
\epsfxsize=4.4truein
\epsfbox[55 32 525 706]{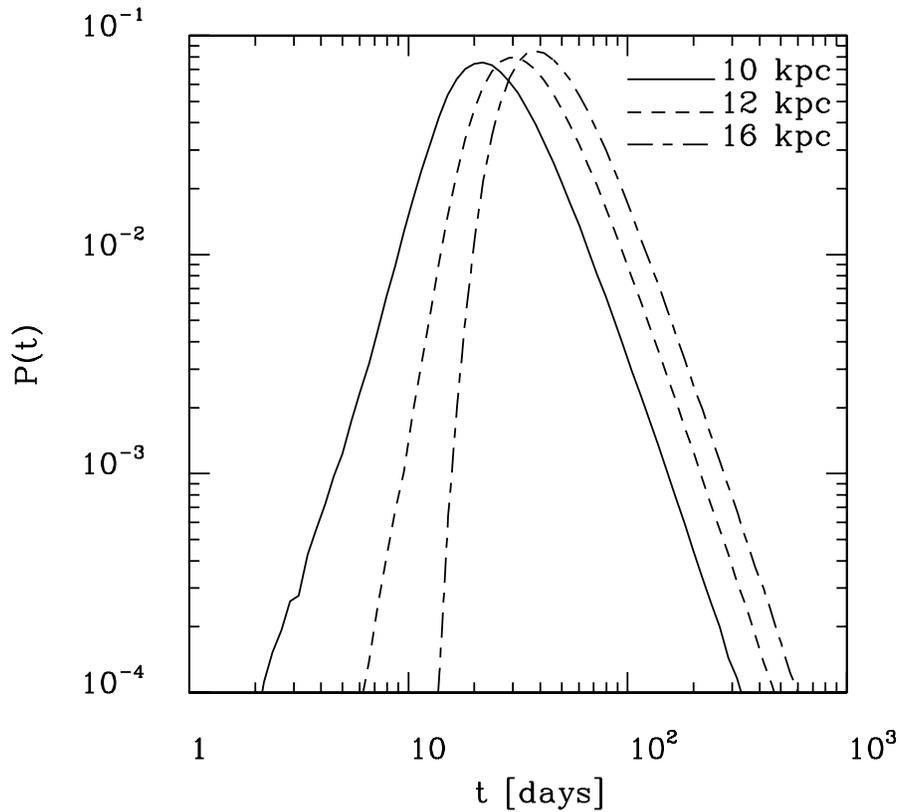}
}
}
\vskip -40pt
\caption{$P(t) \cdot dt$  is the probability, for a fixed source distance,
that a lensing event on a source has an event duration between $t$ and $t+ dt$.
Three cases, a bulge source at $10 \kpc$, and  disk sources at $12$ and 
$16 \kpc$ have been shown. In all three cases, the lens is in the bulge
and has a mass of $1\msun$. Isotropic, Gaussian velocity distributions
were assumed for both the bulge and the disk, with dispersion equal to
$150 \kms$ and $50 \kms$, respectively.  
}
\end{figure}
\vfill\eject

\newpage

\begin{deluxetable}{ccccc}   
\small
\tablewidth{0pt}
\tablenum{1}
\tablecaption{Model 1, $m_t=20$}
\tablehead{
\tablevspace{-.2cm}
\colhead{Model} & 
\colhead{Component (i)} & 
\colhead{$10^6\, \tau_{i} $} & 
\colhead{$100\, f_i$} &
\colhead{$10^8\, \left( {\frac{\tau_{i} \cdot f_{i}}{\tau} } \right)$} \\
\tablevspace{-.5cm}
}
\startdata
\tablevspace{-.2cm}
Unperturbed & Disk foreground & 0.58 & 10.8 & 2.9\\
Unperturbed & Disk background & 4.51 & 6.8&14.1 \\
Unperturbed & Disk Total & 2.085 &17.5& 16.9\\
Unperturbed & Disk $>15 \,\kpc$ & 11.52 &0.2&1.0 \\
Unperturbed & Disk $>11 \,\kpc$ & 8.13 & 1.6 &5.1 \\
Unperturbed & Bulge & 2.18 & 82.5 &83.1\\
Unperturbed & Total & 2.16 &100.0&100.0 \\
\tablevspace{.15cm}

Flare and warp & Disk foreground &0.58&10.8&2.9\\
Flare and warp & Disk background &4.68&6.8&14.7\\
Flare and warp & Disk Total & 2.16&17.6&17.5\\
Flare and warp &Disk $>15 \,\kpc$  &12.88&0.3&1.8\\
Flare and warp &Disk $>11 \,\kpc$  &8.58&1.7&6.8\\
Flare and warp & Bulge & 2.18&82.30&82.5 \\
Flare and warp & Total & 2.18 &100.0&100.0 \\
\tablevspace{.15cm}

SD, $v_{G,\phi}=50$& Disk foreground &0 .60&12.1&3.3\\
SD, $v_{G,\phi}=50$& Disk background & 4.50 & 6.1&12.3\\
SD, $v_{G,\phi}=50$& Disk Total & 1.90&18.2&15.6\\
SD, $v_{G,\phi}=50$& Disk $>15 \,\kpc$ & 11.57&0.15&0.8 \\
SD, $v_{G,\phi}=50$& Disk $>11 \,\kpc$ & 8.21&1.3&5.0 \\
SD, $v_{G,\phi}=50$& Bulge & 2.29 &81.8&84.4 \\
SD, $v_{G,\phi}=50$& Disk+Bulge & 2.22 &100.0&100.0 \\
\tablevspace{.15cm}

SD, $v_{G,\phi}=150$& Disk foreground &0.59&11.5&3.1\\
SD, $v_{G,\phi}=150$& Disk background &4.60&6.4&13.4\\
SD, $v_{G,\phi}=150$& Disk Total & 2.02&17.9&16.5\\
SD, $v_{G,\phi}=150$ & Disk $>15 \,\kpc$&  11.65&0.2&1.2\\
SD, $v_{G,\phi}=150$ & Disk $>11 \,\kpc$&  8.34&1.5&5.9\\
SD, $v_{G,\phi}=150$& Bulge & 2.23&82.1&83.5 \\
SD, $v_{G,\phi}=150$& Disk+Bulge & 2.19 &100.0 &100.0\\
\tablevspace{.15cm}

SD, $v_{G,\phi}=250$& Disk foreground &0.58&11.2&3.0\\
SD, $v_{G,\phi}=250$& Disk background &4.62&6.6&13.95\\
SD, $v_{G,\phi}=250$& Disk Total & 2.08&17.8&16.9\\
SD, $v_{G,\phi}=250$&Disk $>15 \,\kpc$  &11.62&0.2&1.3\\
SD, $v_{G,\phi}=250$&Disk $>11 \,\kpc$  &8.3&1.6&6.3\\
SD, $v_{G,\phi}=250$& Bulge & 2.20&82.2&83.1 \\
SD, $v_{G,\phi}=250$& Disk+Bulge & 2.18 &100.0&100.0 \\

\enddata
\end{deluxetable}

\begin{deluxetable}{ccccc}   
\small
\tablewidth{0pt}
\tablenum{2}
\tablecaption{Model 1, $m_t=22$}
\tablehead{
\tablevspace{-.2cm}
\colhead{Model} & 
\colhead{Component} & 
\colhead{$10^6\, \tau_{i} $} & 
\colhead{$100\, f_{i}$} &
\colhead{$10^8\, \left( {\frac{\tau_{i} \cdot f_{i}}{\tau} } \right)$} \\
\tablevspace{-.5cm}
}
\startdata
\tablevspace{-.2cm}
Unperturbed & Disk foreground & 0.62 & 9.2 & 2.4\\
Unperturbed & Disk background & 5.24 & 8.3&18.0 \\
Unperturbed & Disk Total & 2.81 &17.5& 20.3\\
Unperturbed & Disk $>15 \,\kpc$ & 11.9 &0.66&3.3 \\
Unperturbed & Disk $>11 \,\kpc$ & 8.81 &2.8&10.2 \\
Unperturbed & Bulge & 2.33 & 82.5 &79.7\\
Unperturbed & Total & 2.41 &100.0&100.0 \\
\tablevspace{.15cm}

Flare and warp & Disk foreground &0.62&9.1&2.3\\
Flare and warp & Disk background &5.97&8.9&21.2\\
Flare and warp & Disk Total & 3.25&18.0&23.5\\
Flare and warp &Disk $>15 \,\kpc$  &13.60&1.3&7.1\\
Flare and warp &Disk $>11 \,\kpc$  &10.04&3.4&13.7\\
Flare and warp & Bulge & 2.33&82.0&76.4 \\
Flare and warp & Total & 2.50 &100.0&100.0 \\
\tablevspace{.15cm}

SD, $v_{G,\phi}=50$& Disk foreground &0 .65&10.4&2.7\\
SD, $v_{G,\phi}=50$& Disk background & 5.15 & 7.4&15.4\\
SD, $v_{G,\phi}=50$& Disk Total & 2.52&17.7&18.2\\
SD, $v_{G,\phi}=50$& Disk $>15\,\kpc$ & 11.92&0.5&2.4  \\
SD, $v_{G,\phi}=50$& Disk $>11 \,\kpc$ & 8.81&2.3&8.1  \\
SD, $v_{G,\phi}=50$& Bulge & 2.44 &82.3&82.0 \\
SD, $v_{G,\phi}=50$& Total & 2.45 &100.0&100.0 \\
\tablevspace{.15cm}

SD, $v_{G,\phi}=150$& Disk foreground &0.63&9.8&2.5\\
SD, $v_{G,\phi}=150$& Disk background &5.43&8.0&17.6\\
SD, $v_{G,\phi}=150$& Disk Total & 2.78&17.8&20.2\\
SD, $v_{G,\phi}=150$ & Disk $>15 \,\pc$&  12.05&0.8&3.9\\
SD, $v_{G,\phi}=150$ & Disk $>11 \,\pc$&  9.09&2.9&10.8\\
SD, $v_{G,\phi}=150$& Bulge & 2.39&82.2& 79.8\\
SD, $v_{G,\phi}=150$& Total & 2.46 &100.0 &100.0\\
\tablevspace{.15cm}

SD, $v_{G,\phi}=250$& Disk foreground &0.63&9.53&2.4\\
SD, $v_{G,\phi}=250$& Disk background &5.47&8.27&18.5\\
SD, $v_{G,\phi}=250$& Disk Total & 2.87&17.8&20.9\\
SD, $v_{G,\phi}=250$&Disk $>15 \,\kpc$  &12.03&0.8&4.1\\
SD, $v_{G,\phi}=250$&Disk $>11 \,\kpc$  &9.12&2.75&10.2\\
SD, $v_{G,\phi}=250$& Bulge & 2.26&82.2&79.1 \\
SD, $v_{G,\phi}=250$& Total & 2.45 &100.0&100.0 \\

\enddata
\end{deluxetable}

\begin{deluxetable}{ccccc}   
\small
\tablewidth{0pt}
\tablenum{3}
\tablecaption{Model 2, $m_t=20$}
\tablehead{
\tablevspace{-.2cm}
\colhead{Model} & 
\colhead{Component} & 
\colhead{$10^6\, \tau_{i} $} & 
\colhead{$100\, f_{i}$} &
\colhead{$10^8\, \left( {\frac{\tau_{i} \cdot f_{i}}{\tau} } \right)$} \\
\tablevspace{-.5cm}
}
\startdata
\tablevspace{-.2cm}
Unperturbed & Disk foreground & 0.62 & 13.8 & 3.6\\
Unperturbed & Disk background & 5.24 & 12.3 &27.5 \\
Unperturbed & Disk Total & 2.80 &26.1& 31.2\\
Unperturbed & Disk $>15 \,\kpc$ & 11.96 &1.0&5.0 \\
Unperturbed & Disk $>11 \,\kpc$ & 8.81 &4.1&15.5 \\
Unperturbed & Bulge & 2.18 & 73.9 &68.8\\
Unperturbed & Total & 2.34 &100.0&2.34 \\
\tablevspace{.15cm}

Flare and warp & Disk foreground &0.62&13.4&3.3\\
Flare and warp & Disk background &6.12&13.3&32.8\\
Flare and warp & Disk Total & 3.34&26.9&36.1\\
Flare and warp &Disk $>15 \,\kpc$  &13.89&2.11&11.8\\
Flare and warp &Disk $>11 \,\kpc$  &10.28&5.2&21.6\\
Flare and warp & Bulge & 2.18&73.1&63.9 \\
Flare and warp & Total & 2.49 &100.0&100.0 \\
\tablevspace{.15cm}

\enddata
\end{deluxetable}

\begin{deluxetable}{ccccc}   
\small
\tablewidth{0pt}
\tablenum{4}
\tablecaption{Model 2, $m_t=22$}
\tablehead{
\tablevspace{-.2cm}
\colhead{Model} & 
\colhead{Component} & 
\colhead{$10^6\, \tau_{i} $} & 
\colhead{$100\, f_{i}$} &
\colhead{$10^8\, \left( {\frac{\tau_{i} \cdot f_{i}}{\tau} } \right)$} \\
\tablevspace{-.5cm}
}
\startdata
\tablevspace{-.2cm}
Unperturbed & Disk foreground & 0.63 & 10.5 & 2.7\\
Unperturbed & Disk background & 5.43 & 10.5&22.9 \\
Unperturbed & Disk Total & 3.03 &20.9& 25.6\\
Unperturbed & Disk $>15 \,\kpc$ & 12.03 &1.0&4.8 \\
Unperturbed & Disk $>11 \,\kpc$ & 8.94 &3.8&13.5 \\
Unperturbed & Bulge & 2.33 & 79.1 &74.4\\
Unperturbed & Total & 2.48 &100.0&100.0 \\
\tablevspace{.15cm}

Flare and warp & Disk foreground &0.63&10.3&2.5\\
Flare and warp & Disk background &6.56&11.6&28.8\\
Flare and warp & Disk Total & 3.77&22.0&31.3\\
Flare and warp &Disk $>15 \,\kpc$  &14.06&2.3&12.0\\
Flare and warp &Disk $>11 \,\kpc$  &10.67&5.0&20.2\\
Flare and warp & Bulge & 2.33&78.0&68.6 \\
Flare and warp & Total & 2.65 &100.0&100.0 \\
\tablevspace{.15cm}

\enddata
\end{deluxetable}

\begin{deluxetable}{ccccc}   
\small
\tablewidth{0pt}
\tablenum{5}
\tablecaption{Model 1, $m_t=20,b=-1 \degree,l=1 \degree$}
\tablehead{
\tablevspace{-.2cm}
\colhead{Model} & 
\colhead{Component} & 
\colhead{$10^6\, \tau_{i} $} & 
\colhead{$100\, f_{i}$} &
\colhead{$10^8\, \left( {\frac{\tau_{i} \cdot f_{i}}{\tau} } \right)$} \\
\tablevspace{-.5cm}
}
\startdata
\tablevspace{-.2cm}
Unperturbed & Disk foreground & 1.0 & 12.2 & 2.8\\
Unperturbed & Disk background & 9.37 & 16.2&35.2 \\
Unperturbed & Disk Total & 5.78 &28.35& 38.0\\
Unperturbed & Disk $>15 \,\kpc$ & 22.88 &0.9&4.9 \\
Unperturbed & Disk $>11 \,\kpc$ & 15.99&5.5&20.5 \\
Unperturbed & Bulge & 3.74 & 71.65 &62.0\\
Unperturbed & Total & 4.31 &100.0&100.0 \\
\tablevspace{.15cm}

\enddata
\end{deluxetable}

\begin{deluxetable}{cccc}   
\small
\tablewidth{0pt}
\tablenum{6}
\tablecaption{Present velocity of Sagittarius dwarf}
\tablehead{
\tablevspace{-.3cm}
\colhead{$v_{h,r}$} & 
\colhead{$v_{h,\phi}$} &
\colhead{$v_{G,r}$} &
\colhead{$v_{G,\phi}$} \\
\tablevspace{-.5cm}
}
\startdata
\tablevspace{-.3cm} 
140.0&-220.2&167.6&0.0 \\
140.0&-270.3&165.0&-50.0 \\
140.0&-370.3&160.0&-150.0 \\
140.0&-470.4&154.9&-250.0 \\
\enddata
\end{deluxetable}

\end{document}